\documentclass[aps,prl,twocolumn,groupedaddress,nofootinbib,notitlepage,showpacs,floatfix]{revtex4-1}
\usepackage{graphicx,graphics,epsfig,subfigure,times,bm,bbm,amssymb,amsmath,amsfonts,amsthm,mathrsfs,MnSymbol}
\usepackage[matrix,frame,arrow]{xypic}
\usepackage[pdfstartview=FitH]{hyperref}
\usepackage{pifont}
\hypersetup{
    colorlinks=true,       	
    linkcolor=blue,          	
   citecolor=blue,        
    filecolor=blue,      	
    urlcolor=blue,           	
    runcolor=blue
}
\usepackage[pdftex]{color}
\usepackage{braket}
\usepackage{enumerate}
\usepackage[normalem]{ulem}
\usepackage{subfigure}
\usepackage{overpic}
\usepackage{dsfont}

\usepackage{multirow}
\newcommand{\be}{\begin{equation}}
\newcommand{\ee}{\end{equation}}
\newcommand{\ba}{\begin{eqnarray}}
\newcommand{\ea}{\end{eqnarray}}
\newcommand\tr{{\mbox{Tr\,}}}
\newcommand{\ignore}[1]{}
\newcommand{\avg}[1]{\left< #1 \right>}
\def\ii{\mathrm{i}}
\def\jj{\mathrm{j}}

\begin{document}

\title{Coherence susceptibility as a probe of quantum phase transitions}

\author{Jin-Jun Chen$^{1}$}
\author{Jian Cui$^{2}$}
\email{jian.cui@uni-ulm.de}
\author{Yu-Ran Zhang$^{1}$}
\author{Heng Fan$^{1,3}$}
\email{ hfan@iphy.ac.cn}
\affiliation{ $^{1}$Beijing National Laboratory for Condensed Matter Physics,
Institute of Physics, Chinese Academy of Sciences, Beijing 100190, China}

\affiliation{ $^{2}$Institute for complex quantum systems \& Center for Integrated Quantum Science and Technology (IQST),Universit\"at Ulm, Albert-Einstein-Allee 11, D-89075 Ulm, Germany}

\affiliation{ $^{3}$Collaborative Innovative Center of Quantum Matter, Beijing 100190, China}

\date{\today }

\begin{abstract}

We introduce a coherence susceptibility method, based on the fact that it signals quantum fluctuations, for identifying quantum phase transitions, which are induced by quantum fluctuations. This method requires no prior knowledge of order parameter, and there is no need for careful considerations concerning the choice of a bipartition of the system. It can identify different types of quantum phase transition points exactly. At finite temperatures, where quantum criticality is influenced by thermal fluctuations, our method can pinpoint the temperature frame of quantum criticality, which perfectly coincides with recent experiments.
\end{abstract}

\maketitle

\section{Introduction}

Fluctuations trigger phase transitions manifesting themselves as the sudden change of the system states in the thermodynamic limit \cite{Sachdevbook,QPT,RevModPhys_QPT}. At absolute zero temperature thermal fluctuations cease, leaving quantum fluctuations the only source for the corresponding transitions, which are known as quantum phase transitions (QPTs).  Arising from the Heisenberg uncertainty principle quantum fluctuations underlie the ``quantumness" of the many-body systems.

Different quantum phases of matter had been in general characterized by their symmetries under Landau's symmetry-broken theory before the striking discovery of the (fractional) quantum Hall effect in the 1980s \cite{FQHall}. Since then exotic quantum phases such as topological ordered phases have emerged as one major topic in condensed matter physics and many-body physics.
In modern times, with the development of quantum information sciences a wide variety of characterizations for quantumness have been proposed for the investigation of QPTs including topological phase transitions.
Entanglement was the first and most famous one \cite{AmicoFiniteTentanglement,RMPentanglement, osterloh2002scaling,NielsenEntQPT,EntQPT}. Quantum discord, which measures the quantum correlation between two components of the system, complemented entanglement in certain situations to detect QPTs \cite{discordSarandy,discord}. As another indication of quantumness, i.e., the lack of local convertibility, was also useful in characterizing quantum phases \cite{PhysRevA.85.022338,Cui-Gu,PhysRevB.88.125117,FranchiniPRX}.
Compared with traditional methods based on condensed matter physics, which underlie low-energy effective theories of local order parameters \cite{wanglei}, the quantum information oriented-methods are provided with a common advantage in that they can study QPTs without any knowledge of order parameters in priority. This merit has been validated in particular through the studies of exotic quantum phases with topological orders \cite{Pollmann,PreskillTOP,HammaTOP,KitaevQPT,FranchiniPRX,DiscordTOP}.
Though remarkable success has been achieved, some obvious drawbacks exist, e.g., a careful
consideration of the bipartition of the system is required by definition. For instance, it has been shown in Ref. \cite{Bipartition} that using the entanglement method some OPTs can be detected only by bipartite entanglement between certain two sublattices but not by pairwise concurrence between two nearest-neighbour sites.

Another important topic in studying QPTs is to find out the influence of thermal fluctuations on the quantum criticality of QPTs, since in the real world QPTs have to be observed and manipulated at finite temperatures. It is known that certain QPTs or quantum criticality persists at non-zero temperatures \cite{RevModPhys_QPT,Suzuki2013,Criticality,AmicoFiniteTentanglement,PhaseDiagram_T,PhaseDiagram_Exp}. For instance, recently the temperature frame of quantum criticality of an Ising model QPT has been measured experimentally \cite{PhaseDiagram_Exp}. The quantum criticality has been reported to survive up to temperature $T \sim 0.4 J$ with $J$ being the coupling strength of the Ising model \cite{PhaseDiagram_Exp}. The closest theoretical prediction was $T \sim 0.5 J$ \cite{Criticality}, which is not so desirable.

Both aspects above are related to the coherence of the system.
Indeed, coherence is a better alternative to depict quantumness as it involves no partition of the system.  Coherence has been widely used in the fields of biological physics \cite{biocoherence} and quantum open systems \cite{opencoherence}. It was mainly characterized by the off-diagonal entries of the density operator. The problem was that the many measures of coherence were merely out of physical intuitions. Only recently has a rigorous framework been established \cite{Baumgratz_Coherence} where the measure based on relative entropy and that on the $l_1$ matrix norm were shown as legitimate measures. Much attention has been paid in this direction since then \cite{CoherenceEnt,Frozencoherence,CoherenceinFiniteD,FidelityCoherence}.
The basic idea in Ref. \cite{Baumgratz_Coherence} is to treat coherence under the resource theory, and the central requirement is that coherence should not increase under incoherent operations.
The following simplified form of relative entropy has been proven as a valid measure of coherence for a given basis:
\ba
C({\rho}) &=& S({\rho}_{diag}) - S({\rho}),
\label{eq:measure}
\ea
where $S(\bullet)$ stands for the von Neumann entropy of $\bullet$ and ${\rho}_{diag}$ is obtained from ${\rho}$ by removing all its off-diagonal entries.

On the other hand, QPTs can be formulated as follows. A many-body system with tunable parameter(s) $\lambda$ is described by the Hamiltonian $H(\lambda)$. For each given $\lambda$ the ground state of the system is labeled as $\ket {\Psi_g(\lambda)}$. At QPT point $\lambda_c$ when slightly varying $\lambda$, the ground state drastically changes. Occurring at absolute zero where thermal fluctuation is completely frozen, QPT roots in the quantum fluctuation only, which is induced by the change of $\lambda$. Hence, the non-analyticity of the ground state at phase transition point can be characterized by the singularity of the coherence susceptibility, which is defined as $\chi^{co} \equiv \partial C(\rho)/\partial \lambda$. Here, $\rho$ can be the density operator of the whole system or the reduced density operator of a subsystem. Throughout this paper the derivative in the coherence susceptibility is carried out numerically, whereas the coherence is calculated analytically because of the exact solutions of the corresponding models.

In this paper, we will establish the relative entropy in Eq.(\ref{eq:measure}) as the coherence measure and its susceptibility as a powerful tool to explore QPTs.
The coherence susceptibility method reflects the origin of QPTs, i.e., quantum fluctuations, and it requires no prior knowledge of the order parameters nor a careful consideration of the bipartition. Its simplicity makes it suitable to study QPTs in general, and to investigate the complicated situation of finite-temperature QPTs. We showcase the performance of our method through several models. The temperature frame of quantum criticality pinpointed by our method perfectly matches the experimental result in the Ising model \cite{PhaseDiagram_Exp}.

\section{Results for zero temperature}

We apply the coherence susceptibility method to two spin models with different types of ordinary QPTs and one model with topological QPT. Since the measure of coherence depends on the basis of writing the density matrix,  we fix the computational basis throughout the paper, i.e., the product of eigenbasis of $\sigma^z$, to write the reduced density matrices of the ground states for these models and then substitute $\rho$ in Eq.(\ref{eq:measure}) with them to calculate the coherence. Note that choosing a different basis may change the value of coherence (and its susceptibility) for each state in the phase diagram; however, the singularity stays at the phase transition point. In very rare cases, the coherence may always equal zero for all states by poorly choosing certain basis, where no useful information can be extracted. Such a negative fine-tuned effect can be avoided in general by taking another basis. One also has to pay attention to the size of the reduced density operator. In general the bigger the reduced density operator, the more information it contains, which means a higher chance to probe the QPT but being more difficult to calculate. In practice one can start from the one-site density operator to calculate the coherence. If the result is trivial, meaning that the coherence is always zero and no information can be extracted, one has to increase the size of the density operator until a non-trivial density operator emerges.

The transverse field Ising model (TFIM) in one spatial dimension is the simplest model with a quantum phase transition. Its Hamiltonian reads
\ba
H_I = -J\sum\limits_{\ii=1}\sigma_{\ii}^z\otimes\sigma_{\ii+1}^z
-B\sum\limits_{\ii=1}\sigma_{\ii}^x,
\label{eq:IsingHamiltonian}
\ea
where $\sigma^{x,z}$ are the usual Pauli matrices.
We can choose the free parameter $\lambda:=J/B$ to study the phase diagram.
It can be solved analytically using Jordan-Wigner transformation \cite{Jordan1928,HeisenbergSolution,IsingSolution}. The one-site reduced density operator for the ground state can be recovered from the expectation values of the Pauli matrices as $\rho_1 = \sum_{\alpha=0}^{3} \avg {\sigma^{\alpha}}\sigma^{\alpha}/2$, where $\sigma^{1,2,3} = \sigma^{x,y,z}$, and $\sigma^{0} = I$ is the identity matrix. Here $\avg {\bullet}$ denotes the average of $\bullet$ over the ground state.
The critical point locates at the second order phase transition point $\lambda_c=1$.
For $0<\lambda<1$, the system is gapped, and the ground state is in the paramagnetic phase with vanishing order parameter $\avg {\sigma_z}$. When $\lambda>1$ the system is also gapped with double degeneracy in the energy spectrum. In the thermodynamic limit the symmetry can be broken spontaneously, and the (degenerate) ground states are in the ferromagnetic phase with non-vanishing order parameter $\avg {\sigma_z}$. The difference between the two branches of the degenerate ground states is the opposite sign of the order parameter, and therefore they are labeled as $\ket {0^+}$ and $\ket {0^-}$, respectively. In this phase the system is also possible to stay in an equal mixture of $\ket {0^+}$ and $\ket {0^-}$, i.e., $\rho_{TG} = (\ket {0^+} \bra {0^+} + \ket {0^-} \bra {0^-})/2 $,  which has zero order parameters. This mixture state is known as the ``thermal ground state."
In this simple model we can investigate the one-site reduced density operator for the system in the thermodynamic limit to detect the QPT.
We study the coherence for both the symmetry broken ground state and the thermal ground state.

It can be deduced from  Ref.\cite{NielsenEntQPT} that for $\ket {0^+}$
\begin{eqnarray}
\avg{\sigma^{z}}=\left\{
\begin{aligned}
&0& , \lambda\leq 1, \\
&(1-\lambda^{-2})^{1/8}& , \lambda>1,
\end{aligned}
\right.
\label{Eq:Ising_z}
\end{eqnarray}
\begin{eqnarray}
\avg{\sigma^{x}}=\frac{1}{\pi}\int_0^{\pi}d\phi\frac{1+\lambda\cos\phi}{\sqrt{1+\lambda^2+2\lambda\cos\phi}},
\label{Eq:Ising_x}
\end{eqnarray}
and $\avg{\sigma^{y}} = 0$. For the thermal ground state the only difference is that $\avg{\sigma^{z}}$ is always zero. Inserting these results to Eq.(\ref{eq:measure}) one can obtain the coherence for TFIM in Fig. 1. The main plot shows that the quantum phase transition is detected by the singularity of the coherence for the symmetry-broken ground state, and the inset shows that it is identified by the divergence of the coherence susceptibility for the thermal ground state.

\begin{figure}
\includegraphics[width=\columnwidth]{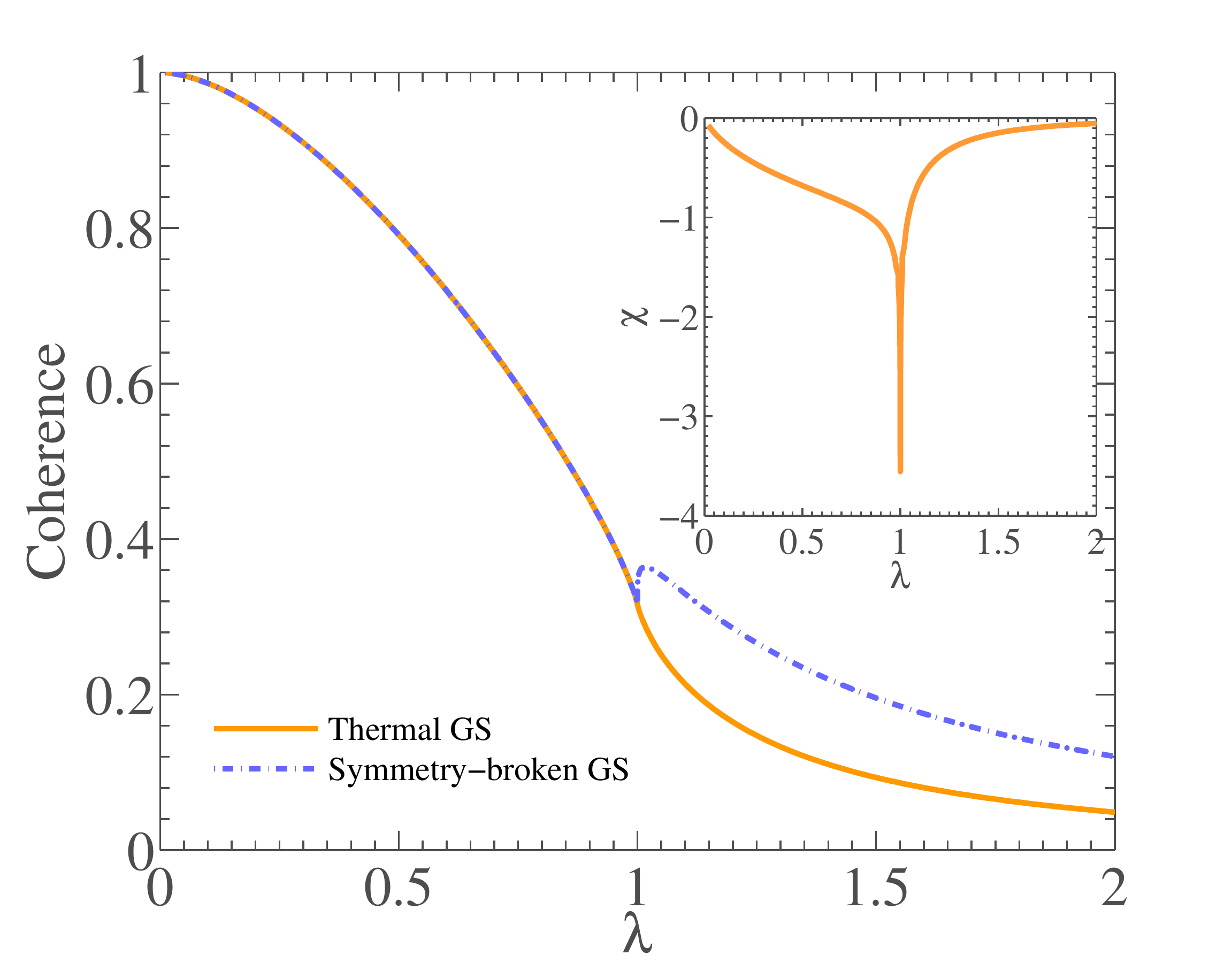}
\caption{Ising model ground state. Main plot: Coherence of one-site reduced density matrix for the symmetry-broken ground state (dash-dot line) and the thermal ground state (solid line). For the symmetry broken state the singularity occurs exactly at the quantum phase transition point. Inset: coherence susceptibility for the thermal ground state. This is obtained by numerically carrying out the derivative of the coherence curve in the main plot with respect to $\lambda$, therefore the actual divergence of $\chi$ at the phase transition point manifests itself as a big negative value in the numerical result due to the finite step.}
\label{fig:Ising_GS}
\end{figure}

A complementary type of quantum phase transition is that in one phase the system is gapped whereas in the whole region of the other phase the system is critical. A spin half $XX$ model is one such example with Hamiltonian
\ba
H_{xx} = -\frac{1}{2}\sum_{\ii=1}^N \bigg[\sigma_{\ii}^x\sigma_{\ii+1}^x + \sigma_{\ii}^y\sigma_{\ii+1}^y \bigg] - \lambda\sum_{\ii=1}^N\sigma_{\ii}^z,
\ea
where the dimensionless parameter $\lambda$ denotes the strength of the external magnetic field in units of the interaction energy.
The phase diagram is symmetric with respect to $\lambda$ \cite{LIEB1961407,XXQPT,SonXX}, therefore we consider only positive $\lambda$. This model can be solved analytically \cite{LIEB1961407,SonXX}. At $\lambda=1$ the system undergoes a first order quantum phase transition.
When $\lambda >1$ the ground state is polarized up for all spins resulting in vanishing coherence.
In the region $0 \le \lambda < 1$ the system is critical and the one-site reduced density operator is diagonal, so its coherence always vanishes according to Eq.(\ref{eq:measure}).
The one-site reduced density operator is trivial in terms of coherence since no information can be extracted.
We can employ the coherence of two adjacent spins to detect the quantum phase transition point.
Its reduced density operator reads $\rho_2 = \sum_{\alpha=0}^{3}\sum_{\beta=0}^{3} \avg {\sigma^{\alpha}\otimes\sigma^{\beta}}\sigma^{\alpha}\otimes\sigma^{\beta}/4$. It has been shown in Refs. \cite{LIEB1961407,SonXX} that the non-trivial non-vanishing coefficients are
$\avg{\sigma^{z}\sigma^{z}} = \big(1-2\arccos(\lambda)/\pi \big)^2 - 4(1-\lambda^2)/\pi^2$, $ \avg{\sigma^{x}\sigma^{x}} = - 2\sin(\arccos(\lambda))/\pi$ and $ \avg{\sigma^z} = 1-2\arccos(\lambda)/\pi$. Substituting these in Eq.(\ref{eq:measure}) one obtains the coherence in Fig. 2, where the singularity at $\lambda=1$ perfectly detects the phase transition.
\begin{figure}
\includegraphics[width=\columnwidth]{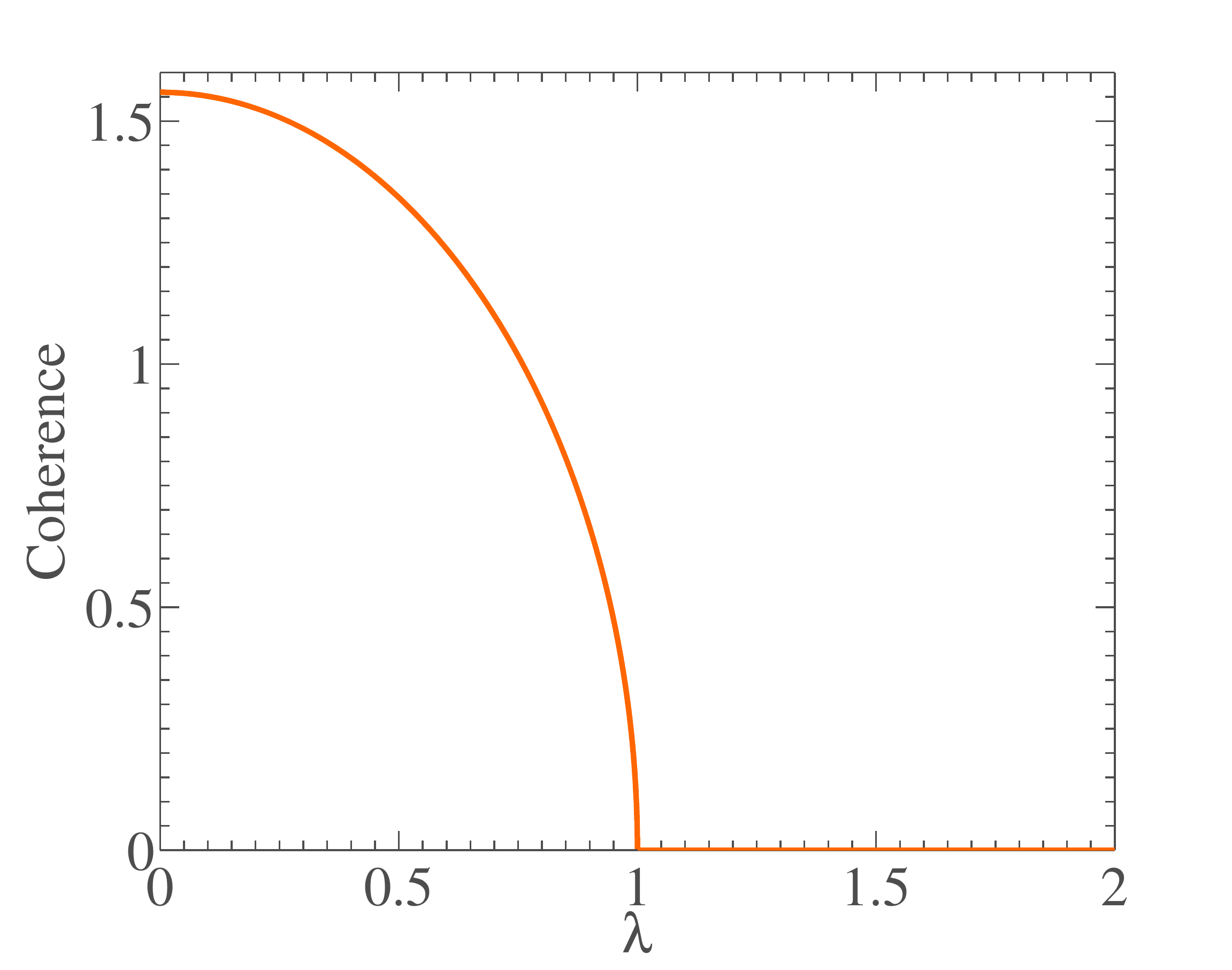}
\caption{The coherence of two adjacent spins for the XX model in the thermodynamic limit. }
\label{fig:xx_c}
\end{figure}

There exists another kind of exotic quantum phase transitions, namely, topological quantum phase transitions, without any local order parameter and therefore they cannot be described by the Landau paradigm \cite{Landau}. We will show through the Kitaev honeycomb model that the topological phase transition can be captured by the coherence susceptibility method.
The Kitaev honeycomb model on a hexagonal lattice with direction-dependent interactions between adjacent lattice sites is an analytically solvable model with topological quantum phase transition \cite{KitaevHoneycomb}. Its Hamiltonian reads
\ba
H_{K} = -\sum_{\alpha=\{x,y,z\}} J_{\alpha}\sum_{(\ii,\jj)\in\alpha-links} \sigma^{\alpha}_\ii\sigma^{\alpha}_\jj,
\ea
where the parameters $J_{\alpha}$ represent the interaction energy between two adjacent spins along the $\alpha$ direction with the $\alpha$ link. We fix $J_x + J_y + J_z = 1$ as the energy unit for this model. The phase diagram is shown in the left inset of Fig. 3. In the shaded areas the system is gapped with Abelian excitation, and in area B it is gapless with non-Abelian excitation.
We take the path $J_y = J_z = (1-J_x)/2$ to study the topological phase transition, which occurs at $J_x=0.5$. The analytical solution of the ground state is applied to derive the reduced density matrix of two adjacent spins with $x$ link in the thermodynamic limit. The only non-vanishing non-trivial coefficient is
$\avg{\sigma^x\sigma^x} = \int_{-\pi}^{\pi} \int_{-\pi}^{\pi}  {\varepsilon}/{\sqrt{\varepsilon^2+\Delta^2}} d\omega_yd\omega_z/{4\pi^2}$ with $\varepsilon = J_x + J_y\cos \omega_y + J_z\cos \omega_z$ and $\Delta = J_y\sin \omega_y + J_z\sin \omega_z$ (see the Appendix for details).
For this model, the coherence susceptibility is defined as $\chi = \partial C/\partial J_x$. Figure 3 depicts the two-site coherence and its susceptibility for the Kitaev model. The topological quantum phase transition at $J_x=0.5$ is well captured as a singular point.
\begin{figure}
\includegraphics[width=\columnwidth]{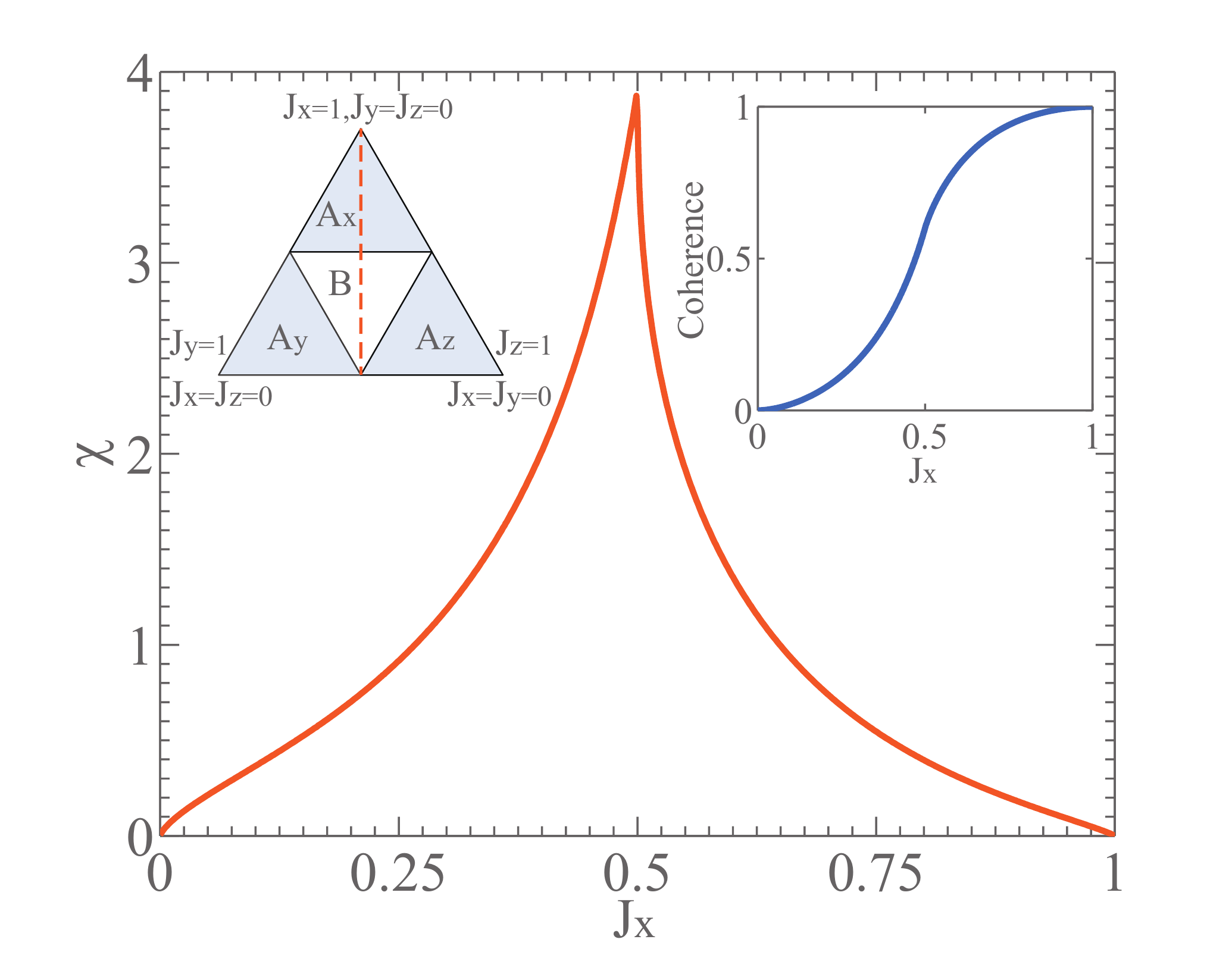}
\caption{Kitaev model ground state. The left inset is the phase diagram of the Kitaev model. We calculate the ground state along the dashed line. The right inset shows the obtained coherence for two sites with an x link. The main figure plots the coherence susceptibility. The singular point occurs exactly at the topological phase transition point $J_x=0.5$. }
\label{fig:Kitaev}
\end{figure}

\section{Results for finite temperatures}

 In real experiments, finite temperature is unavoidable due to the third law of thermodynamics, hence the influence of thermal fluctuation has to be considered.  At finite temperatures, quantum phase transitions are washed out by the thermal fluctuations, nevertheless
certain phase transitions still persist in 2D systems \cite{RevModPhys_QPT}. For 1D systems, e.g., the TFIM, the singularity of the free energy disappears at finite temperatures \cite{Suzuki2013} however, the quantum phase transition has so profound an influence that the criticality is detectable even at finite temperatures \cite{Criticality,AmicoFiniteTentanglement,PhaseDiagram_T}. In other 1D examples, the phase transition points at zero temperature were reported to be signalled by the quantum correlations at finite temperatures \cite{discord}. In this context the system to study is in thermal equilibrium with a reservoir at temperature $T$. Thus, it is described by the Gibbs thermal state as $\rho=\exp(-H/k_BT)/Z$, where $Z=\tr\{ \exp(-H/k_BT)\}$ is the partition function, $k_B$ is the Boltzmann constant, and $k_BT$ sets the energy scale of the thermal fluctuations. Using TFIM [Eq.(\ref{eq:IsingHamiltonian})] at $k_BT$ ranging from $0$ to $1$ as an example, we showcase here that the coherence susceptibility is a powerful tool to study the finite temperature phase diagram.  In order to compare with the previous results in this literature, we rewrite the Hamiltonian as $H_I = -\sum\limits_{\ii=1}^N\sigma_{\ii}^z\otimes\sigma_{\ii+1}^z
-\lambda\sum\limits_{\ii=1}^N\sigma_{\ii}^x$.
The finite temperature phase diagram was predicted in Ref. \cite{PhaseDiagram_T} [also see the left inset of Fig. 4 where the two dotted crossover lines separate it into three distinct regions: renormalized classical (RC), quantum critical (QC), and quantum disordered (QD) region]. On the crossover lines, $k_BT$ equals the energy gap, i.e., $2|\lambda-\lambda_c|$.
Recently this diagram has been confirmed in experiments \cite{PhaseDiagram_Exp}. Moreover, the experiment showed that the criticality could survive up to $k_BT \approx 0.4$.
In the following we show that all the above complex properties can be detected by the simple coherence susceptibility method.

\begin{figure}
\includegraphics[width=\columnwidth]{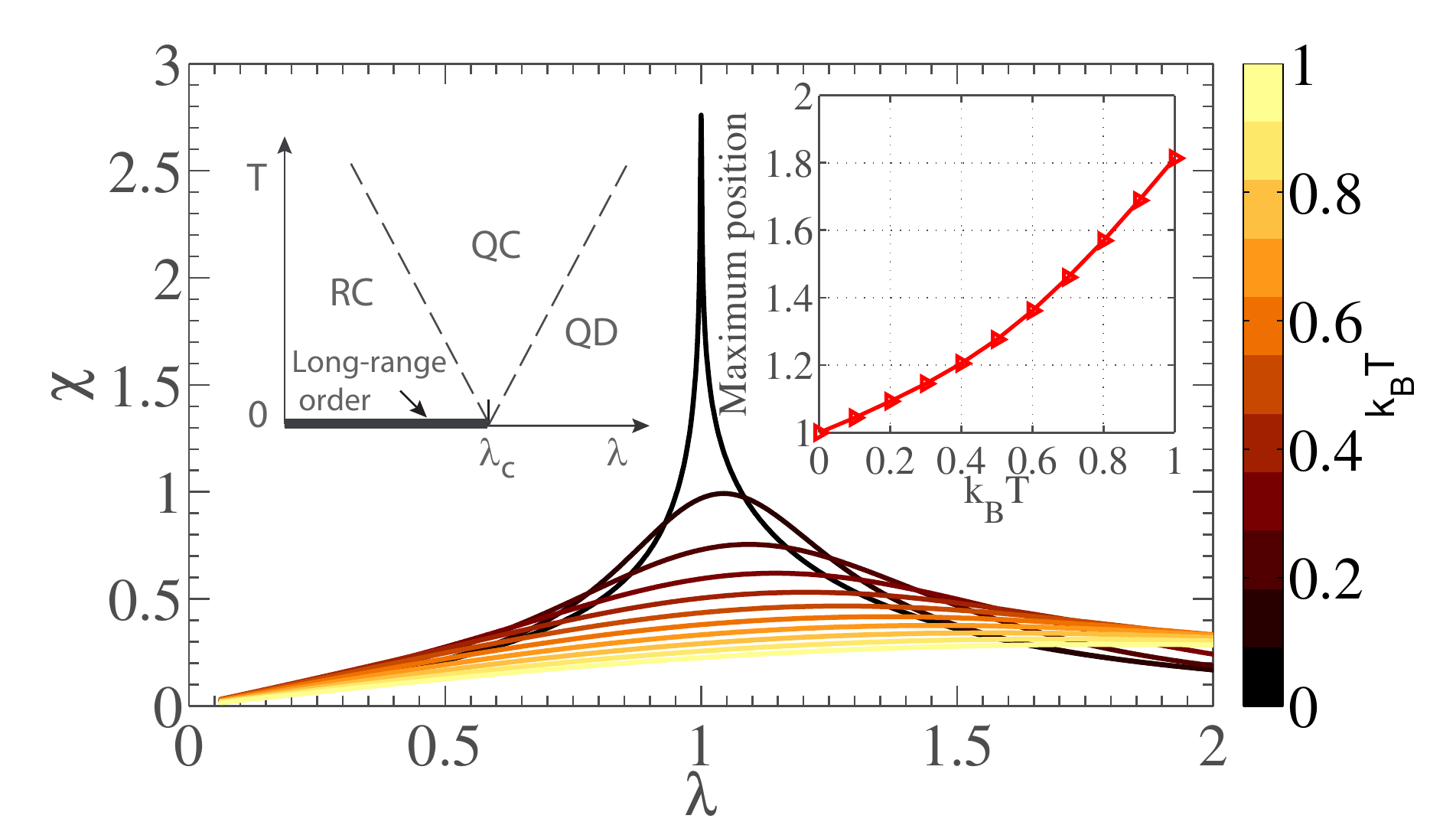}
\caption{Ising model thermal states at different temperatures. Main plot: coherence susceptibility at different temperatures. At finite temperatures the quantum phase transition is smoothed out due to thermal fluctuation. As a result the singularity deforms into maximum points. Left inset: Phase diagram at finite temperatures. Right inset: the locations of the maximum point at different temperatures. Here $k_BT$ is in unit of the interaction energy in the system Hamiltonian.  }
\label{fig:Ising_thermal_chi}
\end{figure}

Again, we take the analytical form of the one-site reduced density operator for the thermal state with
\begin{eqnarray}
\avg{\sigma^{x}}=-\frac{1}{\pi}\int_0^{\pi}d\phi\frac{(\lambda+\cos\phi)\tanh(-\omega_{\phi}/k_BT)}{\omega_{\phi}},
\label{Eq:Ising_thermal_x}
\end{eqnarray}
where $\omega_{\phi}=\sqrt{1+\lambda^2+2\lambda\cos\phi}$, and $\avg{\sigma^{y}} = \avg{\sigma^{z}}=0$ for the thermal state \cite{XY_correlation}.
Fig. 4 demonstrates its coherence susceptibility. One can find that with growing temperature the singular behavior deforms from the divergence to maximum, which shows that the quantum phase transition is smoothed out by the increasing thermal fluctuations. The right inset accounts for the location of the maximum $\chi$, which is labeled as $\lambda_M(T)$, at different temperatures. When $k_BT$ is below 0.4 the $\lambda_M(T)$ is linear with T, which indicates the criticality of the model. The linear scaling changes at $k_BT = 0.4$, which coincides with the temperature frame in experiment \cite{PhaseDiagram_Exp}.
If one compares the $\lambda_M(T)$ curve with the phase diagram in the left inset, it is clear that when $k_BT \le 0.4$, the $\lambda_M(T)$ are located on the second crossover line $k_BT=2(\lambda-\lambda_c)$, so that it not only detects the temperature range of criticality, but also pinpoints the crossover boundary between the QC and QD regions. From the experimental point of view, this feature is useful to extrapolate the quantum phase transition point (occurring at $T=0$) with data from real experiment (at finite temperatures).

To compare, in Ref. \cite{discord} quantum discord was reported to be able to exactly detect the quantum critical point of the $XXZ$ model even at finite temperatures; however, we found that quantum discord is unable to detect the temperature frame for the quantum criticality of Ising model. Since quantum discord is invariant under local unitary transformations, we can change the basis by exchanging $\sigma^x$ and $\sigma^z$ on every site. Then the reduced density matrix of the two nearest neighbor sites can be written as $\rho_2 = [I\otimes I + \braket {\sigma^x\sigma^x} \sigma^z\otimes\sigma^z + \braket {\sigma^y\sigma^y} \sigma^y\otimes\sigma^y + \braket{\sigma^z\sigma^z} \sigma^x\otimes\sigma^x + \braket{\sigma^x} (\sigma^z\otimes I +I\otimes\sigma^z)]/4$, which is of the ``X" type so that the analytical solution for quantum discord in Ref. \cite{QuantumDiscordCalculation} can be applied. $\braket{\sigma^x}$ is already known from Eq.(\ref{Eq:Ising_thermal_x}), and the others can be easily derived from Ref. \cite{XY_correlation} as
\begin{eqnarray}
\avg{\sigma^{x}\sigma^{x}}&=& \avg{\sigma^x} \avg{\sigma^x}- G_{+}G_{-}, \\
\avg{\sigma^{y}\sigma^{y}}&=&G_{+} ,\\
\avg{\sigma^{z}\sigma^{z}}&=&G_{-} ,
\label{Eq:Ising_thermal_xxetc}
\end{eqnarray}
where
\begin{eqnarray}
G_{\pm}&=& -\frac{1}{\pi} \int_0^{\pi}d\phi \frac{(\lambda+\cos\phi)\tanh(-\omega_{\phi}/k_BT)}{\omega_{\phi}}\\ \nonumber
& &\pm \frac{1}{\pi} \int_0^{\pi}d\phi \frac{sin^2\phi \tanh(-\omega_{\phi}/k_BT)}{\omega_{\phi}},
\end{eqnarray}
with $\omega_{\phi}=\sqrt{1+\lambda^2+2\lambda\cos\phi}$.

\begin{figure}
\includegraphics[width=\columnwidth]{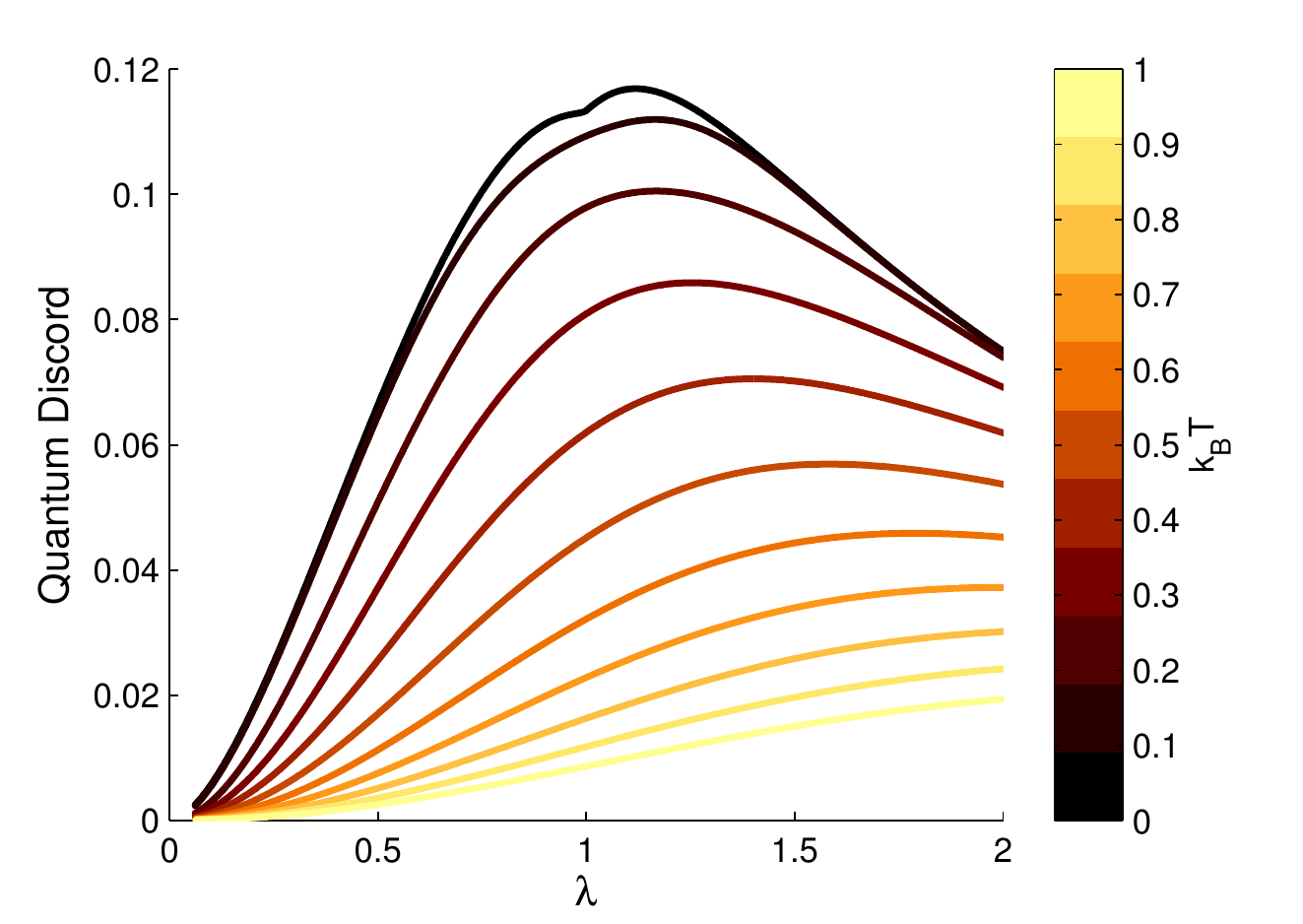}
\caption{Quantum discord for Ising model thermal states at different temperatures. From top to bottom the temperature increases from $k_BT=0$ to 1 in unit of the interaction energy in the system Hamiltonian. }
\label{fig:Ising_thermal_QD}
\end{figure}

\begin{figure}
\includegraphics[width=\columnwidth]{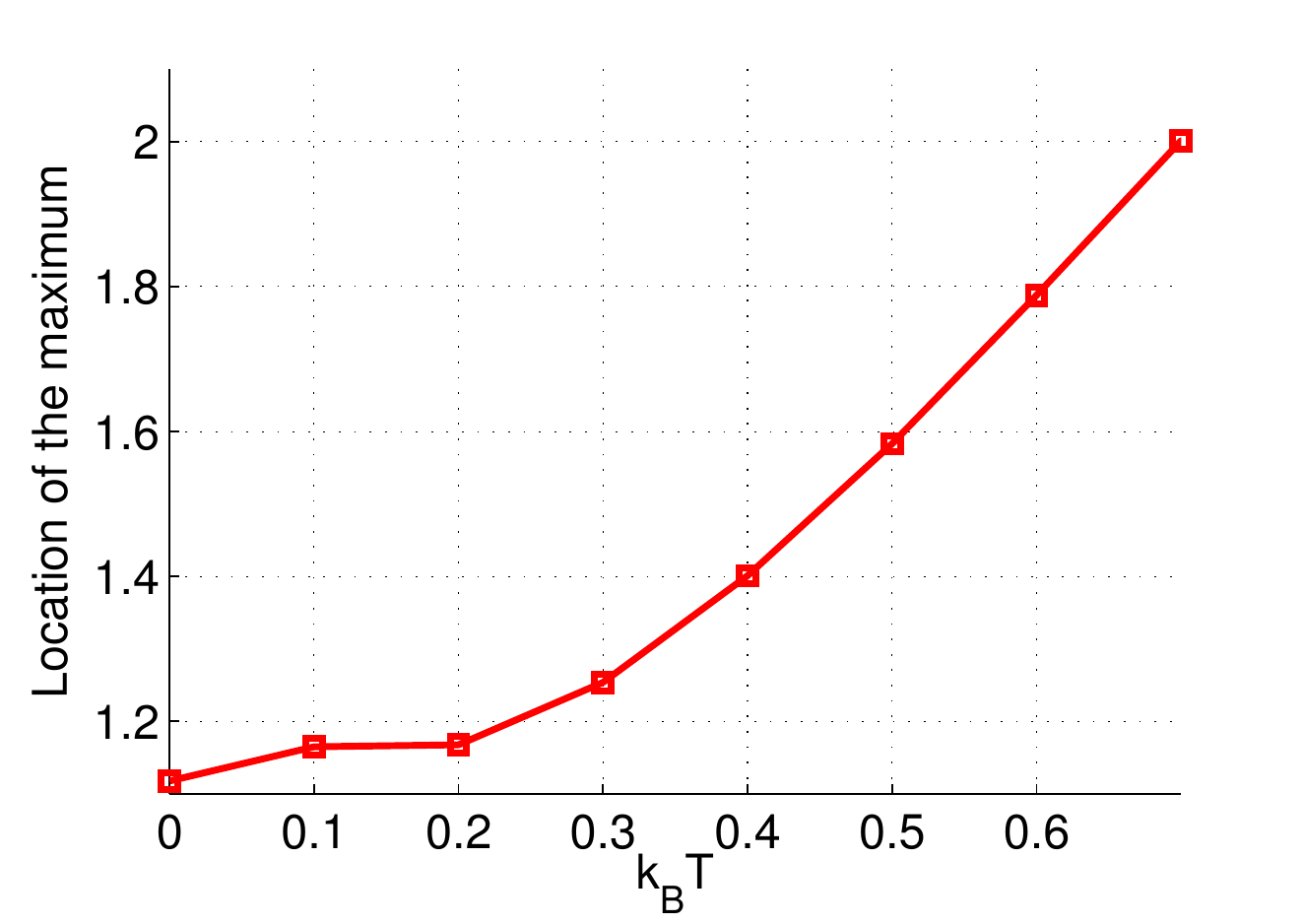}
\caption{ The $\lambda$ location of the maximal quantum discord with respect to temperatures for the Ising model. We show it only up to $k_BT=0.7$ because the real maximal quantum discord for higher temperatures is out of the calculation scope of $\lambda$. Here $k_BT$ is in unit of the interaction energy in the system Hamiltonian. }
\label{fig:Ising_thermal_QD_maximum}
\end{figure}

In Fig. \ref{fig:Ising_thermal_QD} we show the quantum discord of the thermal states at different temperatures. From these curves it is not so clear how the quantum critical point can be detected by quantum discord. In Fig. \ref{fig:Ising_thermal_QD_maximum} we demonstrate the maximal quantum discord as a function of temperature, from which it is not obvious to make any conclusion on the temperature frame of the quantum criticality.
In this example the quantum coherence susceptibility method works better than the quantum discord method.

\section{Discussion}

We have introduced a coherence susceptibility method to detect general quantum phase transitions.
Different from the entanglement method and quantum discord method  \cite{discord}, which succeed in detecting QPTs essentially by selecting various correlation functions in some educated ways, the underlying idea of the present paper is that coherence represents the quantumness of a system, and its susceptibility reflects the quantum fluctuation, which triggers the quantum phase transition. As a result at quantum phase transition points the coherence susceptibility should show singularities. The advantage of this method over other existing methods is that it requires no prior knowledge of the order parameters of the system and no careful consideration of the bipartition is needed. In this sense, it is akin to the fidelity susceptibility method but with a different mechanism as its foundation.
We have applied this method to various quantum systems with continued, first order, and topological phase transitions. In all cases, our method can detect the phase transition points exactly.

The effect of finite temperature on the coherence susceptibility method has also been discussed.
Our method not only can extrapolate the quantum phase transition point with finite temperature data, but also pinpoint the crossover boundary between quantum critical and disordered regions and estimate the temperature frame of quantum criticality. The temperature frame predicted by our method matches the experimental conclusion perfectly, which has filled the gap between the previous theoretical result \cite{Criticality} and the recent experiment.
It is an interesting open question as to whether the crossover boundary that separates the quantum critical region and the renormalized classical region can be detected through the coherence susceptibility of bigger constituent systems.
 One step further would be to generalize the coherence susceptibility method from identifying quantum phase transition points to the quantum-classical crossover\cite{QCcrossover,QCcrossoverIsing}. It is also interesting to investigate the dynamics of closed or open systems at both zero and finite temperatures using the coherence susceptibility.

\section{ Acknowledgement}
We are thankful to Dong Wang, Dr.Pietro Silvi, and Prof.Simone Montangero for fruitful discussions. The research is supported by the NSFC with Grant No. 91536108 and the Chinese Academy of Sciences with Grant No. XDB01010000. J. C. acknowledges the support of the European Commission-funded FET project ``RySQ" with Grant No. 640378. This work is partially supported by MOST of China (Grant No.2016YFA0302104 and No. 2016YFA0300600) and CAS (Grant No.XDB01010000 and No. XDB21030300).

\section{Appendix: The exact solution of the Kitaev honeycomb model}
In this appendix we show how to derive the ground state of the Kitaev model and calculate the correlation function of two sites with an $x$ link using the original method in Ref. \cite{KitaevHoneycomb}.

\begin{figure}[h]
\includegraphics[width=.8\columnwidth]{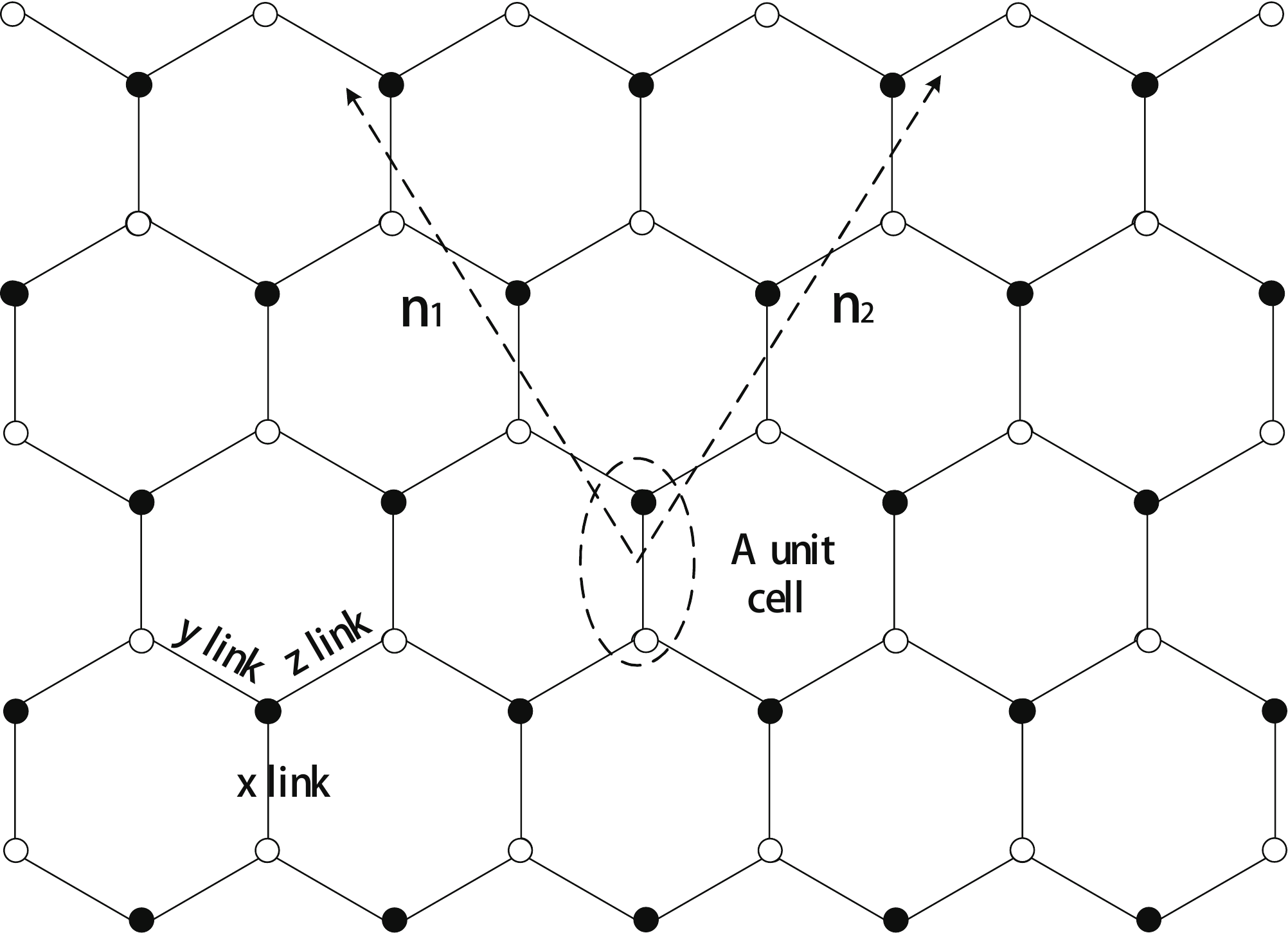}
\caption{A graphic representation of the Kitaev model. There are two sublattices ( empty
and full circles). Each unit cell (marked by elliptic circle) contains  one site of each kind.
Three types of bonds are labeled by ``x-links,'' ``y-links,'' and ``z-links''. We choose the coordinate axes  in $\textbf{n}_1$ and $\textbf{n}_2$
directions.}
\label{fig:KitaevLattice}
\end{figure}

We first introduce the Majorana operators. At each site, we define four Majorana operators $c^{\alpha}$, with $\alpha=0,x,y,z$, satisfying $(c^{\alpha})^{\dagger}=c^{\alpha}$, $\{c^{\alpha},c^{\beta}\}=2\delta_{\alpha\beta}$, and
$c^xc^yc^zc^0=1$. We denote $c_j^0$ by $c_j$ and represent the Pauli operators by the Majorana operators as
\begin{eqnarray}
\sigma_j^a=ic_j^a c_j,  \quad  (a=x,y,z)
\end{eqnarray}
Since the values of $\alpha$ in the Kitaev model are determined by
the site index $j$ and $k$, we can rewrite the Hamiltonian as
\begin{eqnarray}
H=\frac{i}{2}\sum_{j,k}J_{{\alpha}_{j,k}}\hat{u}_{j,k}c_jc_k \quad (\hat{u}_{j,k}\equiv ic_j^{a}c_k^{a})
\end{eqnarray}
Kitaev has showed that $\hat{u}_{j,k}^2=1$, [$\hat{u}_{j,k},H$]=0, and $\hat{u}_{j,k}$ commute with each other. We take $u_{j,k}=1$ for all links because this vortex-free configuration has the lowest energy \cite{Lieb}. We follow the convention that $j$ is the sublattice presented by the empty circles and $k$ is the other sublattice  presented by the full circles in Fig. \ref{fig:KitaevLattice}.

Then, we choose the unit cell containing one empty circle and  one full circle along the $x$ link, shown in the Fig. \ref{fig:KitaevLattice}. We rewrite the site index
$j$ as $(\textbf{s},\lambda)$, where $\textbf{s}$ refers to the location of
the unit cell, and $\lambda$ describes the two different kinds
of sublattices(empty circles
take the value $1$, full circles take the value $2$).
Thus, the Hamiltonian becomes
\begin{eqnarray}
 H&=&\frac{i}{2}\sum_{\textbf{s},\lambda,\textbf{t},\mu}J_{\textbf{s},\lambda,\textbf{t},\mu}c_{\textbf{s},\lambda}c_{\textbf{t},\mu}.
\end{eqnarray}
Actually, $J_{\textbf{s},\lambda,t,\mu}$ is
determined by three indexes $\lambda,\mu,$ and
$\textbf{t}-\textbf{s}$. Then, using the Fourier
transformation, the Hamiltonian takes the form
\begin{eqnarray}
H&=&i\sum_{\textbf{q}}\sum_{\lambda,\mu=1}^2\widetilde{J}_{\lambda,\mu}(\textbf{q})a_{-\textbf{q},\lambda}a_{\textbf{q},\mu}.
\end{eqnarray}
with
\begin{eqnarray}
\widetilde{J}_{\lambda,\mu}(\textbf{q})&=&\sum_\textbf{t}e^{i\textbf{q}\cdot
\textbf{r}_t}J_{0,\lambda;\textbf{t},\mu},\nonumber\\
a_{{\textbf{q}},\lambda}&=&\sqrt{\frac{1}{2L^2}}\sum_\textbf{s}e^{-i\textbf{q}\cdot\textbf{r}_s}c_{\textbf{s},\lambda},
\end{eqnarray}
where $a_{\textbf{q},\lambda}$ satisfies
$a_{-\textbf{q},\lambda}=a_{\textbf{q},\lambda}^{\dag}$,
$a_{\textbf{q},\lambda}^2=0$,
$\{a_{\textbf{p},\lambda},a_{\textbf{q},\mu}^{\dag}\}\equiv
\delta_{\textbf{p}\textbf{q}}\delta_{\lambda,\mu}
$, and other anticommutators are all equal to zero.

After simple calculations we obtain that
$\widetilde{J}_{1,1}(\textbf{q})=\widetilde{J}_{2,2}(\textbf{q})=0$. We choose two directions $\textbf{n}_1=\sqrt{3}/2\textbf{e}_y-\textbf{e}_z/2$, and $\textbf{n}_2=\sqrt{3}/2\textbf{e}_y+\textbf{e}_z/2$, shown in Fig. \ref{fig:KitaevLattice}.  Then,
$\widetilde{J}_{1,2}(\textbf{q})=J_x+J_ye^{i\textbf{q}\cdot\textbf{n}_1}+J_ze^{i\textbf{q}\cdot\textbf{n}_2}$,
$\widetilde{J}_{2,1}(\textbf{q})=-\widetilde{J}_{1,2}^*(\textbf{q})$.
Let $f(\textbf{q})=\varepsilon(\textbf{q})+i\Delta(\textbf{q})$,
and choose $\overrightarrow{q_y}$ to be in the direction of
$\textbf{n}_1$, and $\overrightarrow{q_z}$ to be in the direction of
$\textbf{n}_2$. Then we have
\begin{eqnarray}
\varepsilon(\textbf{q})&=&J_x+J_y\cos q_y+J_z\cos q_z\nonumber,\\
\Delta(\textbf{q})&=&J_y\sin q_y+J_z\sin q_z ,
\end{eqnarray}
where $q_y$ and $q_z$ take values $q_y,q_z=2\pi n/L$,
$n=-(L-1)/2,\cdots,(L-1)/2$.

Next, we use the Bogoliubov transformation to diagonalize the Hamiltonian,
\begin{eqnarray}
C_{\textbf{q},1}&=&u_{\textbf{q}}a_{\textbf{q},1}+v_{\textbf{q}}a_{\textbf{q},2}\nonumber,\\
C_{\textbf{q},1}^{\dag}&=&u_{\textbf{q}}^{*}a_{\textbf{q},1}^{\dag}+v_{\textbf{q}}^{*}a_{\textbf{q},2}^{\dag}\nonumber,\\
C_{\textbf{q},2}&=&v_{\textbf{q}}^{*}a_{\textbf{q},1}-u_{\textbf{q}}^{*}a_{\textbf{q},2}\nonumber,\\
C_{\textbf{q},2}^{\dag}&=&v_{\textbf{q}}a^{\dag}_{\textbf{q},1}-u_{\textbf{q}}a^{\dag}_{\textbf{q},2},
\end{eqnarray}
where, $u_\textbf{q}=1/\sqrt{2}$,
$v_{\textbf{q}}=if_{\textbf{q}}/(\sqrt{2}|f_{\textbf{q}}|)$,
$v_{-\textbf{q}}=-v_{\textbf{q}}^*$, and the new operators satisfy
$\{C_{\textbf{q},\lambda},C_{\textbf{p},\mu}^{\dag}\}=\delta_{\textbf{p}\textbf{q}}\delta_{\lambda,\mu}$, $C_{\textbf{q},\lambda}^2=0$,
$C_{-\textbf{q},1}=-2u_{\textbf{q}}^*v_{\textbf{q}}^*C_{\textbf{q},2}^{\dag}$,
and $C_{\textbf{q},1}^{\dag}C_{\textbf{q},1}=1-C_{-\textbf{q},2}^{\dag}C_{-\textbf{q},2}$.
Then, the Hamiltonian reads
\begin{eqnarray}
H=\sum_{\textbf{q}}|f_{\textbf{q}}|(1-2C_{\textbf{q},2}^{\dag}C_{\textbf{q},2}).
\end{eqnarray}
The normalized ground state is
\begin{eqnarray}
|G\rangle=\prod_{\textbf{q}}C_{\textbf{q},2}^{\dag}|0\rangle,
\end{eqnarray}
 with $C_{\textbf{q},2}|0\rangle=0$.
The energy gap is $2\min_{\textbf{q}}\{|f_{\textbf{q}}|\}$.

\section*{Correlation functions and Reduced density matrix of the Kitaev model}

The two-site reduced density matrix is the joint state of two spins at sites $i$ and $j$ and takes the form $\rho_{i,j}=\Sigma_{\alpha,\beta=0}^3\langle \sigma_{i}^{\alpha}\sigma_{j}^{\beta}\rangle\sigma_{i}^{\alpha}\sigma_{j}^{\beta}/4$, where $\sigma^{0}$ is the identity,  and $\sigma^{1,2,3}=\sigma^{x,y,z}$. So we need 16 correlation functions to construct the two-site density matrix.
We calculate the correlation functions of two nearest lattices linked by an $x$ bond, and it is obvious that $\langle\sigma_{\textbf{r},1}^{0}\sigma_{\textbf{r},2}^{0}\rangle=1$. For the rest of the correlation functions, only $\langle\sigma_{\textbf{r},1}^{x}\sigma_{\textbf{r},2}^{x}\rangle$ is nonzero. Thus, the two-site reduced density matrix along the $x$ link has the form,
\begin{equation}
\rho_2=  \frac{1}{4}
\left(
  \begin{array}{cccc}
   1 & 0 & 0 & \langle\sigma^x\sigma^x\rangle\\
   0 &1&\langle\sigma^x\sigma^x\rangle & 0\\
   0 & \langle\sigma^x\sigma^x\rangle & 1 & 0\\
   \langle\sigma^x\sigma^x\rangle & 0 & 0 & 1\\
  \end{array}
\right)
\end{equation}
Then, we calculate the correlation function $\langle\sigma_{\textbf{r},1}^{x}\sigma_{\textbf{r},2}^{x}\rangle$  using the analytic result of the model:

\begin{eqnarray}
\langle\sigma^x_{\textbf{r},1}\sigma^x_{\textbf{r},2}\rangle&=&\langle b^x_{\textbf{r},1}b^x_{\textbf{r},2}\frac{2}{L^2}\sum_{\textbf{q},\textbf{q}^{'}}e^{i(\textbf{q}+\textbf{q}^{'})\cdot\textbf{r}}a_{\textbf{q},1}a_{\textbf{q}^{'},2}\rangle\nonumber\\
&=&-i\frac{2}{L^2}\sum_{\textbf{q},\textbf{q}^{'}}e^{i(\textbf{q}+\textbf{q}^{'})\cdot\textbf{r}}\langle
a_{\textbf{q},1}a_{\textbf{q}^{'},2}\rangle\nonumber.
\end{eqnarray}
By using the relation
\begin{eqnarray}
\langle a_{\textbf{q},1}a_{\textbf{q}^{'},2}\rangle&=&\langle(u^*_{\textbf{q}}C_{\textbf{q},1}+v_{\textbf{q}}C_{\textbf{q},2})(v^*_{\textbf{q}^{'}}C_{\textbf{q}^{'},1}-u_{\textbf{q}^{'}}C_{\textbf{q}^{'},2})\rangle\nonumber\\
&=&-u^*_{\textbf{q}}u_{\textbf{q}^{'}}\langle C_{\textbf{q},1}C_{\textbf{q}^{'},2}\rangle\nonumber\\
&=&\frac{i}{2}\delta_{\textbf{q},-\textbf{q}^{'}}\frac{f_{\textbf{q}}}{|f_{\textbf{q}}|}\nonumber,
\end{eqnarray}
we obtain the correlation function
\begin{eqnarray}
\langle\sigma^x_{\textbf{r},1}\sigma^x_{\textbf{r},2}\rangle&=&\frac{1}{L^2}\sum_{{\textbf{q}}}\frac{f_{\textbf{q}}}{|f_{\textbf{q}}|}=\frac{1}{2L^2}\sum_{{\textbf{q}}}\frac{f_{\textbf{q}}+f_{-\textbf{q}}}{|f_{\textbf{q}}|}\nonumber\\
&=&\frac{1}{L^2}\sum_{\textbf{q}}\frac{\varepsilon_{\textbf{q}}}{E_{\textbf{q}}},
\end{eqnarray}
where
$E_{\textbf{q}}=|f_{\textbf{q}}|=\sqrt{\varepsilon^2_{\textbf{q}}+\Delta^2_{\textbf{q}}}$.

In the thermodynamic limit, we use the  continuous $\omega_{y(z)}$  replacing $q_{y(z)}$, then  the correlation function has the form

\begin{eqnarray}
\langle\sigma^x\sigma^x\rangle&=&\frac{1}{4\pi^2}\!\!\int_{-\pi}^{\pi}\int_{-\pi}^{\pi}\frac{\varepsilon}{\sqrt{\varepsilon^2+\Delta^2}}\;d\omega_yd\omega_z,
\end{eqnarray}
with $\varepsilon=J_x+J_y\cos \omega_y+J_z\cos \omega_z$ and $\Delta=J_y\sin \omega_y+J_z\sin_z$.

\bibliography{CoherenceCriticalreference}

\clearpage

\end{document}